\documentclass[runningheads]{llncs}
\usepackage[utf8]{inputenc}
\usepackage[T1]{fontenc}
\usepackage{times}
\usepackage{graphicx}
\usepackage{amsmath}
\usepackage{amssymb}
\usepackage{amsmath}
\usepackage{amssymb}
\usepackage{amsfonts}
\usepackage{stmaryrd}
\usepackage{esvect}
\usepackage{xspace}
\usepackage{listings}
\usepackage{algorithm}
\usepackage[noend]{algpseudocode}
\usepackage{multicol}
\usepackage{xcolor,sectsty}
\usepackage{tabularx}
\usepackage{multirow}
\usepackage{array}
\usepackage{bm}  

\newcolumntype{Y}{>{\centering\arraybackslash}X}
\makeatletter
\newcommand{\verbatimfont}[1]{\renewcommand{\verbatim@font}{\ttfamily#1}}
\makeatother
\usepackage{bussproofs}
\usepackage{hyperref}
\usepackage{bsymb}
\usepackage{relsize}
\usepackage{array}
\usepackage{multirow}

\def\BibTeX{{\rm B\kern-.05em{\sc i\kern-.025em b}\kern-.08em
    T\kern-.1667em\lower.7ex\hbox{E}\kern-.125emX}}
\newcommand{\DEF}
    {\mathrel{\stackrel{\scriptscriptstyle\mathchar"7001}{=}}}

\newcounter{cntbib}
{
  \begin{list}%
  {[#1\arabic{cntbib}]}
  {\usecounter{cntbib}
   \setlength{\rightmargin}{0in}
   \setlength{\listparindent}{0in}
   \settowidth{\labelwidth}{* [#199]}
   \setlength{\leftmargin}{0in}
   \addtolength{\leftmargin}{\labelsep}
   \addtolength{\leftmargin}{\labelwidth}
  }
}
{\end{list}}

\newcommand{\comment}[1]{}

\newcounter{adressecnt}

\newcommand{\myrule}[3]{\raise#1 \hbox{\vrule width#2 height#3 depth0ex}}

\newcommand{\dleq}%
  {\mathrel{\mbox{$\sqsubseteq$\hspace{-.55em}%
  \myrule{.22ex}{.04em}{1.16ex}\hspace{.52em}}}}

\newcommand{\dgeq}%
  {\mathrel{\mbox{$\sqsupseteq$\hspace{-.29em}%
  \myrule{.16ex}{.04em}{1.28ex}\hspace{.52em}}}}

\newcommand{\djoin}%
  {\mathbin{\mbox{\rlap{$\sqcup$}\hspace{.1em}%
  \myrule{.32ex}{.47em}{.08ex}\hspace{.09em}}}}
\newcommand{\ddbigsqcup}%
  {\mbox{\rlap{$\displaystyle\bigsqcup$}\hspace{1.3pt}%
  \myrule{.3pt}{8.5pt}{.8pt}\hspace{.13em}}}
\newcommand{\dtbigsqcup}%
  {\mbox{\rlap{$\bigsqcup$}\hspace{1.2pt}%
  \myrule{.6pt}{5.9pt}{.6pt}\hspace{.11em}}}
\newcommand{\dbigsqcup}%
  {\mathop{\mathchoice{\ddbigsqcup}{\dtbigsqcup}{\dtbigsqcup}{\dtbigsqcup}}}

\newcommand{\dmeet}%
  {\mathbin{\mbox{\rlap{$\sqcap$}\hspace{.1em}%
  \myrule{1ex}{.47em}{.08ex}\hspace{.09em}}}}

\newcommand{\ddbigsqcap}%
  {\mbox{\hspace{.06em}%
  \myrule{-4pt}{.8pt}{14pt}\rlap{\myrule{9.24pt}{8.4pt}{.8pt}}%
  \myrule{5.2pt}{8.4pt}{.8pt}\myrule{-4pt}{.8pt}{14pt}\hspace{.08em}}}
\newcommand{\dtbigsqcap}%
  {\mbox{\hspace{.04em}\myrule{-2pt}{.6pt}{9.9pt}%
  \rlap{\myrule{7.3pt}{5.9pt}{.6pt}}%
  \myrule{4.7pt}{5.9pt}{.6pt}\myrule{-2pt}{.6pt}{9.9pt}\hspace{.05em}}}

\newcounter{rulenbr}
\newcommand{\nbrule}[2]{
    \refstepcounter{rulenbr}\label{#2}\mbox{#1-\therulenbr}}

\newcounter{ionbr}

\newcounter{mflistnbr}

\newlength{\labelsp}

\newlength{\esppar}
\newlength{\parthlength}
\newlength{\mfemlength}
\newcommand{\m}{\hspace*{1em}}
\newcommand{\mm}{\hspace*{2em}}

\newlength{\pgmindent}
\newlength{\pgmtabindent}
{\small \begin{center}{\bf Résumé}\end{center}\begin{quote}\indent}
{\end{quote}}
\reversemarginpar

\usepackage{latexsym}
\newlength{\symbwidth}

\font\msx=msam10
\font\msy=msbm10

\newfam\msxfam \textfont\msxfam=\msx
\newfam\msyfam \textfont\msyfam=\msy

\def\famletter#1{\ifcase #1 0\or 1\or 2\or 3\or 4\or 5\or 6\or 7\or
        8\or 9\or A\or B\or C\or D\or E\or F\fi}

\edef\fx{\famletter\msxfam}

\def\@myop#1{\mathop{\mathstrut{#1}}\nolimits}

\def\_{\leavevmode \vbox{\hrule width0.5em}}

\let\xforall=\forall
\let\xexists=\exists
\let\xlambda=\lambda

\def\p#1{\mathrel{\ooalign{\hfil$\mapstochar\mkern 5mu$\hfil\cr$#1$}}}

\let\mc=\mathchardef

\def    \pow            {\mbox{${\cal P}$}}
\def    \po1            {\mbox{${\cal P}_1$}}

\def    \lambda         {\@myop{\xlambda}}

\def    \lor            {\mathrel{\vee}}

\def    \forall         {\@myop{\xforall}}
\def    \exists         {\@myop{\xexists}}

\def    \dom            {\mbox{\textsf{dom}}}
\def    \ran            {\mbox{\textsf{ran}}}
\def    \id            {\mbox{\textsf{id}}}
\def    \comp           {\mathbin{\raise
                        0.6ex\hbox{\oalign{\hfil$\scriptscriptstyle
                        \rm o$\hfil\cr\hfil$\scriptscriptstyle\rm 9$\hfil}}}}

\let    \fun            \rightarrow
\def    \pfun           {\p\fun}
\mc     \inj            "3\fx1A

\mc     \surj           "3\fx10
\def    \na1            {\mbox{${\cal N}_1$}}

\def    \int            {\mbox{${\cal Z}$}}

\def    \finse1         {\mbox{${\cal F}_1$}}

\def    \seq            {\@myop{{\rm seq}}}
\def    \cat            {\mathbin{\raise 0.8ex\hbox{$\mathchar"2\fx61$}}}

\newcommand{\FLO}{\ensuremath{\downpitchfork}\xspace}

\newcommand{\ASTD}{\textsf{\small ASTD}}

\newcommand{\qflowconst}{\textsf{\small Qflow}}

\newcommand{\Envk}{\textsf{\small Env}}

\newcommand{\qflowk}{\mbox{$\FLO{:}$}}

\newcommand{\stdec}{\raisebox{-0.25ex}{$\scriptstyle \circ$}}

\newcommand{\qflowStk}{\qflowk\stdec}

\newcommand{\recordk}[1]{\langle#1\rangle}

\newcommand{\booleank}{\textsf{\small Boolean}}
\newcommand{\eventk}{\textsf{\small Event}}

\newcommand{\namek}{\textsf{\small Name}}
\newcommand{\statek}{\textsf{\small State}}

\newcommand{\termk}{\textsf{\small Term}}
\newcommand{\Vark}{\textsf{\small Var}}

\newcommand{\initk}{\mathit{init}}
\newcommand{\isFinalk}{\mathit{final}}

\usepackage{lmodern, mathtools}

\DeclareFontFamily{U}  {MnSymbolA}{}
\DeclareSymbolFont{MnSyA}         {U}  {MnSymbolA}{m}{n}
\SetSymbolFont{MnSyA}       {bold}{U}  {MnSymbolA}{b}{n}
\DeclareFontShape{U}{MnSymbolA}{m}{n}{
<-6>  MnSymbolA5
<6-7>  MnSymbolA6
<7-8>  MnSymbolA7
<8-9>  MnSymbolA8
<9-10> MnSymbolA9
<10-12> MnSymbolA10
<12->   MnSymbolA12}{}
\DeclareFontShape{U}{MnSymbolA}{b}{n}{
<-6>  MnSymbolA-Bold5
<6-7>  MnSymbolA-Bold6
<7-8>  MnSymbolA-Bold7
<8-9>  MnSymbolA-Bold8
<9-10> MnSymbolA-Bold9
<10-12> MnSymbolA-Bold10
<12->   MnSymbolA-Bold12}{}

\makeatletter
\newcommand{\xleftfork}[2][]{%
\ext@arrow 0079\xleftforkfill@{#1}{#2}%
}
\newcommand{\xleftforkfill@}{%
\arrowfill@\Mnrelbar\Mnrelbar{\mathrel{\leftpitchfork}}
}
\newcommand{\xrightfork}[2][]{%
\ext@arrow 0097\xrightforkfill@{#1}{#2}%
}

\DeclareMathSymbol{\rightpitchfork}{\mathrel}{MnSyA}{"88}%
\DeclareMathSymbol{\leftpitchfork}{\mathrel}{MnSyA}{"8A}%
\DeclareMathSymbol{\Mnrelbar}{\mathrel}{MnSyA}{"D0}%
\newcommand{\downpitchfork}{\mathrel{\scalebox{1.2}{\rotatebox[origin=c]{270}{${\rightpitchfork}$}}}}

\usepackage{scalerel}

\newcommand{\forked}{\mathbin{\ThisStyle{{\supset}\kern-\dimexpr.5\LMex+3pt\relax{-}}}}

\newcommand{\LionelToDo}[1]{}

\usepackage{caption}

\begin{document}
\title{ASTD Patterns for Integrated Continuous Anomaly Detection In Data Logs}
%
%
\author{Chaymae El Jabri\inst{1}\orcidID{0009-0002-7933-8874} \and
Marc Frappier\inst{1}\orcidID{0000-0002-4402-2514} \and
Pierre-Martin Tardif\inst{1}\orcidID{0000-0002-7413-6897}}
\authorrunning{C. El Jabri et al.}
%
\institute{Université de Sherbrooke, Sherbrooke, Canada }
\maketitle              
\begin{abstract}
This paper investigates the use of the ASTD language for ensemble anomaly detection in data logs. It uses a sliding window technique for continuous learning in data streams, coupled with updating learning models upon the completion of each window to maintain accurate detection and align with current data trends. It proposes ASTD patterns for combining learning models, especially in the context of unsupervised learning, which is commonly used for data streams. To facilitate this, a new ASTD operator is proposed, the Quantified Flow, which enables the seamless combination of learning models while ensuring that the specification remains concise. Our contribution is a specification pattern, highlighting the capacity of ASTDs to abstract and modularize anomaly detection systems. The ASTD language provides a unique approach to develop data flow anomaly detection systems, grounded in the combination of processes through the graphical representation of the language operators. This simplifies the design task for developers, who can focus primarily on defining the functional operations that constitute the system.

\keywords{ASTD\and Anomaly detection \and Continuous learning.}
\end{abstract}
\section{Introduction}
In today's digital age, protecting IT infrastructure from cyberattacks and security breaches is critical for organizations to ensure daily operations, store sensitive data and manage customer information. Anomaly detection techniques can help organizations identify unusual patterns and behaviors in their systems so they can respond quickly and prevent potential security incidents. Anomaly detection techniques are instrumental in diverse areas, including fraud detection, network security, and intrusion detection within business applications~\cite{b1}. 

Recognizing the pivotal role of anomaly detection systems in ensuring the security and reliability of various applications, from cybersecurity to industrial monitoring, it is crucial to acknowledge the challenges associated with their development~\cite{b2}. Effectively addressing these challenges becomes imperative for successfully deploying robust and adaptive detection systems.

In the realm of anomaly detection systems, a formidable challenge arises from the dynamic nature of data patterns. To maintain the system's accuracy over time, periodic model re-training becomes imperative. Nils Baumann et al.~\cite{b3} underscore the critical importance of automating the re-training process to adapt to evolving data patterns seamlessly. This challenge necessitates implementing robust mechanisms that detect anomalies and autonomously refine their understanding of normal and abnormal behaviors in the ever-changing data landscape.

The intricacy of learning systems poses yet another significant challenge, encompassing multifaceted phases such as data pre-processing and model training. Benjamin Benni et al.~\cite{b4} delve into a comprehensive analysis of this complexity, shedding light on the intricate processes that form the backbone of effective anomaly detection. Addressing this challenge requires the development of streamlined strategies to simplify the various phases, ensuring that the learning system can efficiently navigate the intricacies of data preprocessing and model training. Overcoming this hurdle is crucial for enhancing anomaly detection systems' overall effectiveness and efficiency.

As detection systems scale up to handle vast amounts of data, a distinct challenge emerges in maintaining modularity to ensure scalability and ease of maintenance. The development of large-scale detection systems demands a careful balance to prevent unwieldy complexity. Baldwin and Clark~\cite{b5} stress the significance of modularity in such systems, emphasizing its pivotal role in facilitating scalability and simplifying maintenance efforts. Successfully addressing this challenge involves designing detection systems with modular architectures that can seamlessly adapt to the increasing demands of data volume and computational resources, ensuring both scalability and ease of long-term maintenance.

 This article introduces a method for developing anomaly detection systems using a specification language called Algebraic State Transition Diagram (ASTD) \cite{b6}. It investigates the extent to which this language reduces the complexity of the detection system by adding an abstraction layer. Additionally, it examines how the graphical representation of the language's operators contributes to easing development efforts by managing the scheduling of various processes within the detection system. ASTD is a graphical and executable notation for composing state machines, offering modularity and flexibility in system development~\cite{b7}. The paper's contributions include (1) The extension of the ASTD language by the Quantified Flow operator to allow the combination of an arbitrary number of models while keeping the specification compact, and (2) The definition of an ASTD specification that represents a pattern on which to base the development of more complex systems; this specification has the following features: - Automated re-training of learning models, - Composition of a set of learning models to detect anomalies in data logs, - Combination of the decisions of each model for each event. The intent is to provide an illustrative example of specifications of anomaly detection systems that can be easily adapted for other contexts or learning methods.

The paper is divided into six sections. In Section~\ref{related_work}, we emphasize the importance of automating the renewal of the learning model in the context of dynamic data, the role of abstraction in reducing system complexity, and modularity, which facilitates maintenance and extension without introducing errors. Section~\ref{section1} introduces the Quantified Flow operator as an extension of the ASTD language to easily combine an arbitrary number of learning models. In Section~\ref{section2}, we present a case study on the detection of unexpected events, implementing the following essential features for unsupervised anomaly detection: - Automation of retraining of learning models using the Sliding Window technique. - Model composition using the Quantified Flow operator. - Decision combination of models through Majority Voting. In section~\ref{experiment}, we assess the performance of the specification in detecting unexpected events during a day of activity, while highlighting the effect of training data renewal and the combination of unsupervised models. Finally, in Sections~\ref{discussion} and~\ref{conclusion}, we summarize our findings and conclude.

\section{Related Work}\label{related_work}

The MLOps~\cite{b8} approach presents a set of principles aimed at standardizing the deployment, management, and monitoring processes of machine learning models in production environments. This approach integrates best practices and tools to optimize the model lifecycle, ensuring their effectiveness and robustness throughout their usage. In this work, we propose a development framework for an unsupervised anomaly detection pipeline, based on statistical learning models. This framework aims to incorporate features that align with certain technical aspects of MLOps, such as:
\begin{enumerate}
    \item \textbf{Periodic Learning}: Regular updating of data using a sliding window approach~\cite{b9,b10}. Gamma~\cite{b9} suggests that in most cases, we are primarily concerned with computing statistics for the recent past rather than the entire history. The sliding window method is useful in this regard as it allows us to focus on the relevant data. There are various window models, including the sequence-based model where the window size is determined by the number of observations, and the timestamp-based model where the window size is determined by duration.
    \item \textbf{Metadata Storage}: Storage of intermediate results associated with the learning model.
    \item \textbf{Entity-Based Data Processing}: Separation and processing of data based on specific entities, such as users or machines, to customize analyses and anomaly detection.
\end{enumerate}
By implementing these features, we aim to create a flexible pipeline that optimizes the development lifecycle of anomaly detection models.

The development of the Anomaly Detection System using ASTD language follows the Model-Driven Engineering (MDE) paradigm, which provides a potential solution to reduce complexities through abstraction~\cite{b11}. MDE advocates for using software models at different levels of abstraction to (semi-) automatically construct software systems. Models serve as abstractions of complex entities; they conceal unwanted details so that modelers can easily focus on their areas of interest. The ASTD language ensures, through the graphical representation of its operators, the combination and scheduling of processes, enabling a focus on the core operations of the detection system.

Modularity in software design is crucial for enhancing maintainability and extensibility by breaking down complex systems into smaller, independent modules. This approach simplifies debugging and maintenance, as changes can be made to individual modules without affecting the entire system. It also promotes code reuse, saving time and effort. Studies such as ~\cite{b11} and ~\cite{b12} emphasize modularity's importance for building scalable and maintainable software systems. The ASTD language embodies this principle, with its specifications organized in a tree structure where each branch represents a specific functionality of the system.

\section{Extended ASTD Formalism}\label{section1}

The Algebraic State Transition Diagram (ASTD)~\cite{b6} is a specification language for modeling and integrating complex systems by extending traditional state machines with process algebra operators. To enable anomaly detection models to work concurrently and independently within an ensemble system, we propose the Quantified Flow operator as an extension of the standard ASTD Flow operator.

\begin{figure}[h!]
    \centerline{\includegraphics[width = \linewidth,height = 4cm]{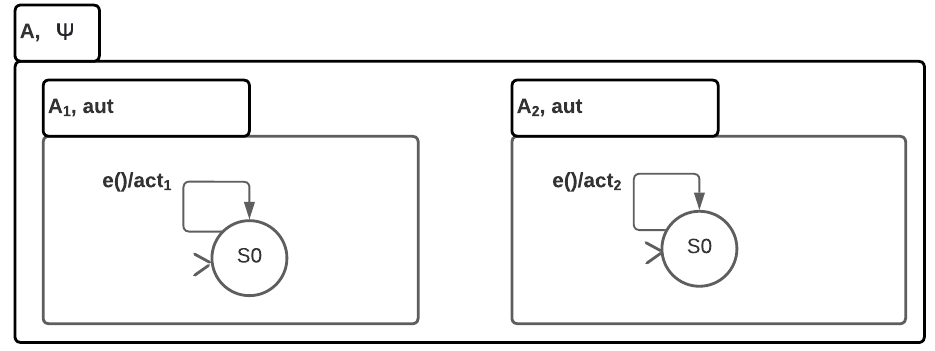}}
    \caption{Flow operator}
    \label{flow_op}
\end{figure}

The Flow operator is an operator similar to the AND state in statecharts. In~\cite{b13}, the flow operator is used for combining the processes of data pre-processing, training, and detection during the development of anomaly detection systems by ASTDs. This is possible thanks to the fact that a single input event can be processed differently in each of the sub-ASTDs of the flow operator according to two distinct actions. This is represented in Figure \ref{flow_op}: When event \textsf{e()} is received, \textsf{act\textsubscript{1}} is executed, followed by \textsf{act\textsubscript{2}}.

The Quantified Flow was introduced to enable support for combining learning models, which involve two phases: training and detection. This operator allows the independent invocation of methods for each phase within the models.

In anomaly detection systems, managing both the training and detection phases across multiple models is important. These models typically perform two main tasks: \textbf{training} and \textbf{evaluation}. During training, the model is updated with new data to improve its performance, while in the evaluation phase, it processes new input to identify anomalies.

\begin{figure}[h!]
    \centerline{\includegraphics[width = \linewidth, height = 4cm]{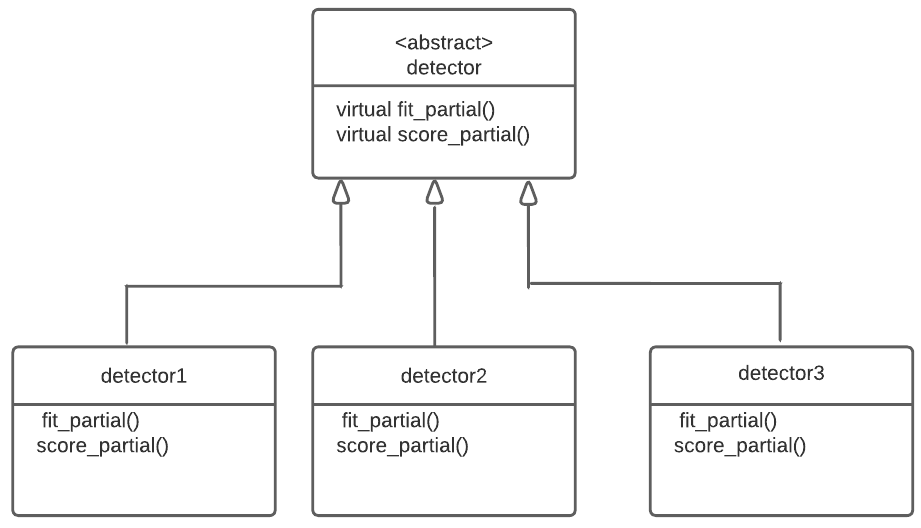}}
    \caption{Class hierarchy for anomaly detection using multiple techniques}
    \label{temp_meth_patt}
\end{figure}

Each task can be executed independently and in parallel across models. To achieve this, we define an abstract structure that captures the general functions of anomaly detection models. In object-oriented programming, this is done by specifying the skeleton of detection models in an abstract class, with methods for training and scoring implemented in subclasses tailored to each model (see Fig.~\ref{temp_meth_patt}).

In order to effectively combine a set of heterogeneous anomaly detection models while maintaining the abstraction of their functioning in an ASTD, we use the abstract class \emph{detector} (shown in Figure~\ref{temp_meth_patt}), which encapsulates the general functions of the detection models. This class defines two primary methods: \textsf{fit\_partial()}: Trains the model incrementally by incorporating each new instance of data, \textsf{score\_partial()}: Returns the score of the input data based on the reference model, indicating whether it represents an anomaly.

\begin{figure}[h!]
    \centerline{\includegraphics[width = \linewidth,height = 4.5cm]{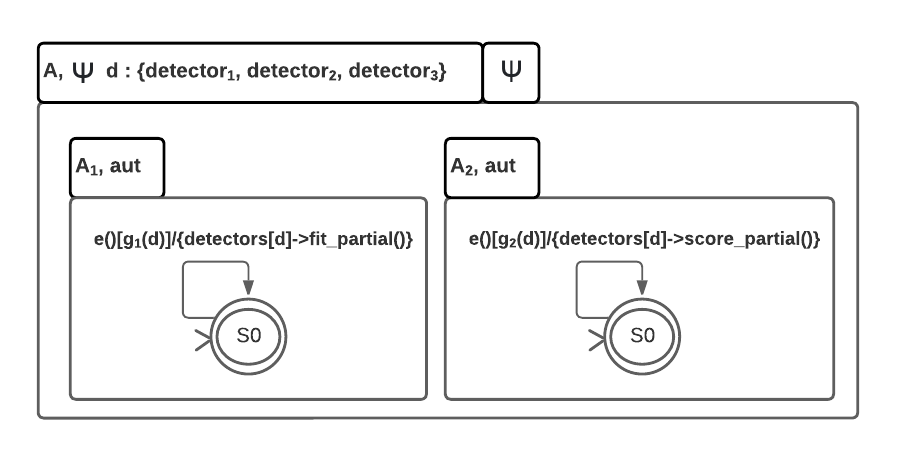}}
    \caption{An ASTD pattern for combining three anomaly detection models using the quantified flow}
    \label{quant_flow_op}
\end{figure}

In Figure~\ref{quant_flow_op}, we present the specification for combining multiple detectors using the Quantified Flow operator. In this example, the attribute \textsf{detectors} is associated with the quantified flow ASTD \textsf{A}, and it is of type \textsf{map<string, detector*>}. This map holds references to instances of the detection models, associating each model's name with its class: 
\{\textsf{'detector1': new detector1(), 'detector2': new detector2(), 'detector3': new detector3()}\}.
Upon receiving an event \textsf{e()}, the Quantified Flow operator uses the quantification variable \textsf{d} to traverse the set of detectors \{\textsf{detector1, detector2, detector3}\}. For each detector, two key actions are sequentially executed, each with its corresponding guard condition: $g_1(d)$ checks if training for model $d$ can begin, and $g_2(d)$ verifies if model $d$ is fully computed and ready to perform detection. The actions involves:
\begin{itemize}
    \item \textsf{fit\_partial()}: The model is trained with the new instance of data.
    \item \textsf{score\_partial()}: The model evaluates the data and returns an anomaly score.
\end{itemize}

This allows each detector to perform its operations independently, without waiting for the others. The Quantified Flow operator facilitates this parallelism while maintaining the abstraction of each model’s function. The use of the \textsf{detector} class ensures that each model behaves according to its unique characteristics, yet all are managed through a unified interface. This structure allows the Quantified Flow operator to handle any learning model—such as \textsf{k-means}, \textsf{KDE}, and \textsf{LOF}—in an integrated and scalable way. 

The example in Figure~\ref{quant_flow_op} illustrates how the hierarchical nature of ASTDs integrates seamlessly with object-oriented class hierarchies. The \textsf{detector} abstract class is utilized to declare the \textsf{detectors} map in the quantified ASTD \textsf{A}, where each instance of \textsf{A} can be associated with its specific type of detector. This design promotes reusability and modularity, ensuring that the system can easily incorporate additional detectors or modify existing ones by simply altering the map and the class instantiations. To generalize to any set of detectors, we must first define classes that inherit from the \emph{detector} interface, implementing the  \textsf{fit\_partial()} and \textsf{score\_partial()} methods. Next, pass a JSON configuration file as a parameter to the specification, containing the list of detector names and the constructors initializing each detector. This will enable dynamic loading of the set of detectors.
\paragraph{Syntax} The quantified flow ASTD subtype has the following structure:
\[
\qflowconst \DEF \recordk{\qflowk,x,T,b}
\]

where \( x \in \Vark \) denotes a quantified variable that can be accessed in read-only mode. The type of this variable is represented by \( T \). \( b \in \ASTD\) refers to the body of the flow, which represents the ASTD that will be executed for each instance of the quantified variable. $\ASTD$ is the abstract type that identifies all the shared characteristics of all ASTD types, $\ASTD \DEF{} \recordk{n,P,V,A_{astd}}$ where $n \in \namek$ is the name of the ASTD, $P$ is a list of parameters, $V$ is a list of attributes, $A_{astd}$ is an action.

Each $\ASTD$ has a set of states, with $\statek$ representing all states. Final states are determined by the function $\isFinalk: \statek \rightarrow \booleank$, and $\initk: \ASTD \rightarrow \statek$ returns the initial state. For a quantified flow, the state is of type $\recordk{\qflowStk, E, f}$, where $\qflowStk$ is the constructor, $E$ is the attribute set, and $f: T \rightarrow \statek$ maps elements $x$ of $T$ to states of $b$, with each state corresponding to an instance of the quantified flow.

Initial and final states are defined as follows. Let $a$ be a quantified flow ASTD:
\[
\begin{array}{@{}rcl@{}}
\initk(a) & \DEF & (\qflowStk,a.E_{init},T \times \{ \initk(a.b)\}) \\
\isFinalk(a,(\qflowStk,E,f)) & \DEF & \forall c : T \cdot \isFinalk(a.b,f(c))
\end{array}
\]

\paragraph{Semantics} The semantics of an ASTD consists of a labeled transition system (LTS),  computed based on the inference rules of ASTD operators~\cite{b14}, which is a subset of  $\statek \times \eventk \times \statek$ representing a set of transitions of the form $ s \xrightarrow{\sigma}_{a} s'$. It means that ASTD $a$ can execute event $\sigma$ from state $s$ and move to state $s'$. The semantics of a nested ASTD depends on the variables declared in its enclosing ASTDs; we use \emph{environments} to represent the values of these variables and the values of ASTD parameters.
An environment is a function of $\Envk \DEF \Vark \pfun \termk$ which assigns values to variables. We need to introduce an auxiliary transition relation that handles environments: $ s \xrightarrow{\sigma, E_e, E'_e}_a s'$,
where environments $E_e,E'_e$ denote the before and after values of variables in the ASTDs enclosing ASTD $a$.

Rule $\qflowk_1$ describes the execution of an event in the quantified flow ASTD. The rule applies when a transition occurs in the body of the ASTD.

\[
\Theta
\;\;\DEF\;\;
\left(
E_g = E_e \ovl E
\m
a.A_{astd}(E''_g,E'_g)
\m
E'_{e} = E_{e} \ovl (V \domsub E'_g)
\m
E' = V \domres E'_g
\right)
\]

where environments $E_e$, $E'_{e}$ denote the before and after values of variables in the ASTDs enclosing ASTD $a$. The ASTD action $A_{astd}$ defines the computation of $E'_g$ from $E''_g$. $E'_{e}$ and $E'$ are extracted by partitioning $E'_g$ using $V$, the set of attributes. \(\Theta\) defines the transformation of environments during a sub-ASTD transition execution.

\begin{center}
\AxiomC{$
\mm
\Omega_{qflow}
\mm
\Theta
$}
\LeftLabel{$\qflowk_{1}$}
\UnaryInfC{$(\qflowStk,E,f)
             \xrightarrow{\sigma,E_e,E'_e}_a
             (\qflowStk,E',f') $}
\DisplayProof
\end{center}
In this context, $(\qflowStk,E,f)$ represents the current state of the quantified flow, where $E$ denotes the environment and $f$ is the function that maps elements of $T$ to the state of the body ASTD $b$. The event $\sigma$ triggers the transition, which changes the enclosing environment from $E_e$  to $E'_e$. After this transition, the function is updated to $f'$, reflecting the changes in the state of the body ASTD. The action $A_{astd}$ governs the changes in the global environment $E'_g$.

We use the following abbreviation to indicate that an ASTD cannot execute a transition from a state $s$ and global attributes $E_g$:
\[
s \not\xrightarrow{\sigma,E_g}_a \;\;\DEF\;\; \neg\exists {\scriptstyle E'_g},s' \cdot s \xrightarrow{\sigma,E_g,E'_g}_a s'
\]
This notation expresses that no transition exists from state $s$ under the event $\sigma$ with global attributes $E_g$.

Premiss $\Omega_{qflow}$ non-deterministically selects a permutation $p$ of $T$ (noted $p \in \pi(T)$) and a sequence of environments $Es$, which store the intermediate results of the computation of $E''_g$ from $E_g$ by iterating over the elements $p(i)$ of $p$ and executing the instances of the quantified flow. The execution order of the instances is chosen non-deterministically.

If the specifier wants deterministic results for the values of attributes, they must ensure that the actions of the instances are commutative. Let $k=|T|$ (the size of $T$):
\[
\Omega_{qflow} \DEF
\left(
\begin{array}{l}
p \in \pi(T) \wedge Es \in 0..k \rightarrow \Envk \wedge Es(0) = E_g \wedge Es(k) = E''_g \wedge \\
\forall i \in 1..k \cdot  
\left(
\begin{array}{l}
f(p(i)) \xrightarrow{\sigma,Es(i-1)\ovl\{x \mapsto p(i)\}, Es(i)}_{a.b} f'(p(i))  \\
\lor \quad \\
Es(i) = Es(i-1) \wedge f(p(i)) \not\xrightarrow{\sigma,Es(i)}_{a.b} 
\end{array}
\right)
\end{array}
\right)
\]
This expression defines the non-deterministic execution of the quantified flow instances, where $p$ is a permutation of $T$, and $Es$ is a sequence of environments that capture intermediate states of the system as each element $p(i)$ is processed.

The existing synchronization operators are not suitable for this application. Quantified synchronization, for instance, allows the parallel execution of its sub-instances and synchronizes their actions based on a set of events called \(\Delta\). The events in \(\Delta\) are executed only when all sub-instances are able to perform them. If learning models are synchronized during the training and detection phases, all models must train simultaneously. This prevents adapting the training of each model to specific conditions, unless these conditions are implemented at the action level rather than the model level. However, this approach limits the extensibility and modularity of the specification: any modification to the quantification set would also require changes to the action code. With Quantified Flow, modifications are localized: only the quantification set needs to be adjusted. Quantified interleave, on the other hand, is a special case of quantified synchronization where the \(\Delta\) set is empty (\(\Delta = \emptyset\)), allowing only one of the sub-instances to execute an event at a time. In this case, there is more independence than necessary to trigger the training or detection of each model, as it requires calling the event \( e(x) \), where \( x \) is the name of the model one wishes to use, for each model individually. In contrast, with Quantified Flow, the independence is optimized: by simply calling the event \( e \), it is executed automatically for all models capable of processing it at the given moment.

ASTDs are supported by the tools eASTD and cASTD~\cite{b15}. eASTD is a graphical editor of ASTD specifications. cASTD is a compiler that translates ASTD specifications into executable code. It first generates an implementation in an abstract, intermediate, imperative language that can be translated into an equivalent executable imperative language like C++, Java, Python. Currently, C++ is the sole translation implemented. The generated code can read the data continuously from a data source, and apply the operations contained in the specification in the order that has been defined thanks to the process algebra operators.

\section{ASTD Specification For Combining Anomaly Detection Models}\label{section2}

In this section, we present a generic ASTD specification for combining a set of heterogeneous detectors. For this purpose, we introduce a real application case, in which we will determine all the components and elements of the specification. The complete specification is found in~\cite{b16}. The main goal of this application is to identify unusual or unexpected events within user activities. These "unexpected events" typically manifest as activities occurring at times when a user is not usually active. Our example is based on the time of occurrence of an event, for illustrative purposes and the sake of simplicity.  Other criteria, or more general techniques for identifying anomalies, could easily be used with our ASTD pattern. For our example, we select three attributes from those available in the log files which are:
\begin{itemize}
    \item Id: uniquely identifies each event, designated in the specification by \textsf{eventId}. 
    \item CreationTime: determines the date and time in Coordinated Universal Time (UTC) when the user performed the activity, designated in the specification by \textsf{eventDate}.
    \item UserId: the user who performed the action 
\end{itemize}

\begin{figure}[h!]
    \centerline{\includegraphics[width = \linewidth,height = 8cm]{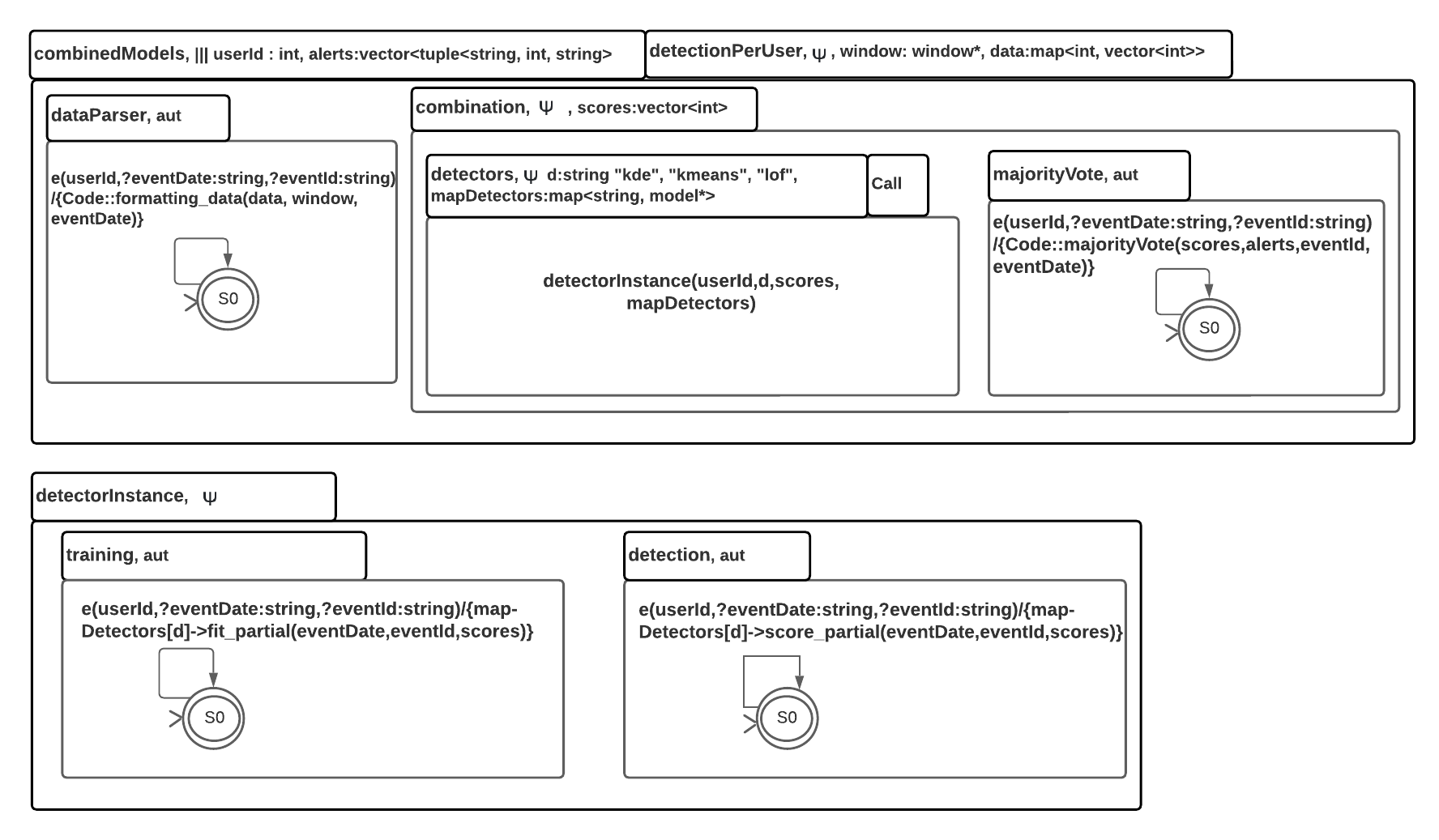}}
    \caption{Specification pattern}
    \label{spec_pattern}
\end{figure} 

\paragraph{Anomaly detection models} We utilize three heterogeneous learning models, which are:
\begin{itemize}
    \item \textsf{K-means}: a clustering algorithm for batch learning adapted to circular data. The number of clusters used is optimized using the silhouette coefficient. The distance used for clustering refers to the time interval between two events occurring at different hours, denoted as $a$ and $b$. The formula to compute this distance is as follows:
    \begin{equation}
    \mathrm{distance}(a,b) =
    \begin{cases}
    \min(b-a, a-b+24), & \text{if}\ a<b \\
    \min(a-b, b-a+24), & \text{otherwise}
    \end{cases}
    \end{equation}
    \item \textsf{KDE}: is a non-parametric statistical technique for estimating the probability density of a random variable. 
    \item \textsf{LOF}: is an unsupervised anomaly detection method that calculates the local density deviation of a given data point from its neighbors. It considers as outliers the samples whose density is significantly lower than that of their neighbors.
\end{itemize}
The data used during the training of the different models consists of the hours of events performed by a user within a day, which means the data is circular in an interval of [0,24[. 

\paragraph{ASTD specification}In Figure~\ref{spec_pattern}, we present the top-level ASTD named \textsf{combinedModels} of type quantified interleave, denoted by $\interleave$ in the upper left tab. It declares a quantification variable \textsf{userId} of type int with an $UnboundedDomain$, which allows the processing of all the users received without the need to identify them a priori. The quantified interleave allows each user to be treated independently by associating an instance of the flow sub-ASTD for each user. The flow combines two sub-ASTDs \textsf{dataParser}, and \textsf{combination}



It has the following attributes: 
\begin{itemize}
    \item \textsf{window} of type \textsf{window*} initialized by \textsf{new window(window\_parameters)}.
    \item \textsf{data} of type $map\langle int,vector \langle int \rangle\rangle$. In cases where the period type is either 'week' or 'day', the keys of the map represent the period number. However, if the type of window is 'instance', the map contains a single key with a value of '0'. The values in the map are the minutes of the occurrence of the events, stored in a vector. 
    \item \textsf{alerts}, which is of type $vector\langle tuple\langle string, int, string\rangle\rangle$. The \textsf{alerts} attribute stores information about abnormal events, including the event identifier, the number of models that flagged the event, and the date of its occurrence.
\end{itemize}  

Additionally, the following parameters are defined:
\begin{itemize}
    \item \textsf{window\_parameters} of type json that respects the following structure $\{windo-w\_size: int, sliding\_size:int, type \in \{'day','week','instance'\}\}$
    \item \textsf{kde\_parameter}: double; the value of the k-percentile that will determine the threshold of probability densities below which an event is considered abnormal. It takes values in [0.5, 5].
    \item \textsf{kmeans\_parameter}: double; the threshold to which the absolute value of the events cluster's z-score is compared. It takes values in [1.5, 2.5].
    \item \textsf{lof\_parameter}: double; the value of the k-percentile that will determine the threshold of LOF scores for the training data, above which any score from the test data is considered abnormal. It takes values in [75, 95].
\end{itemize}

The sub-ASTD named \textsf{detectionPerUser} is of type flow denoted by $\qflowk$, which is a binary operator. It allows the same event to be treated by its two sub-ASTDs, and the combination of the latter two by sharing inherited variables. The right sub-ASTD is \textsf{DataParser} of automaton type, consisting of a single initial and final state (S0) having a loop transition labeled with the event pattern \textsf{e(userId, ?eventDate: string, ?eventId: string)} and the action \textsf{formatting\_data(data, window, eventDate)}, which adds each received event to the training set and defines the data belonging to the current window according to the type of period chosen as shown in the algorithm~\ref{formatting_data}, The methods $add\_instance$ and $add\_period$ of the window class can be found in the window.cpp file at ~\cite{b17}. 

\begin{algorithm}[h!]
\caption{formatting\_data}
\label{formatting_data}
\begin{algorithmic}[1]
\State \textbf{Input:} $data$, $window$, $eventDate$
\State \textbf{Output:} $data$, $window$ updated
\State int $hour = \text{get\_hour}(\text{CreationDate})$
\State string $type = \text{window} \rightarrow \text{getType}()$
\If{$type$ == "day" or $type$ == "week"}
    \State $period = \text{Compute\_period}(\text{CreationDate}, type)$
    \State add $minute$ to $data[period]$
    \State $\text{std::vector}\langle \text{int} \rangle \text{periodsToDelete} = \text{window} \rightarrow \text{add\_period}(period)$
    \State delete the periods in $periodsToDelete$ from the map $data$
\EndIf
\If{$type$ == "instance"}
    \State add $minute$ to $data[0]$
    \State bool $sliding\_on = \text{window} \rightarrow \text{add\_instance}(minute)$
    \State int $sliding\_size = \text{window} \rightarrow \text{getSliding\_size}()$
    \If{$sliding\_on$}
        \State $data[0] \gets$ delete elements from $data[0]$ from start to $sliding\_size$
    \EndIf
\EndIf
\end{algorithmic}
\end{algorithm}

The left sub-ASTD is named \textsf{Combination} it is a flow with the parameter: \textsf{scores} of type $vector \langle int\rangle$. It stores the scores of value 0 or 1 of an input data for each detection model. At the level of the left ASTD referred to as \textsf{detectors}, we establish an attribute called $mapDetectors$, which is of type $map\langle string,model*\rangle$. Here, the term "model" represents an abstract class from which three distinct learning models inherit: namely, k-means, kernel density estimation (KDE), and the local outlier factor (LOF). $\textsf{mapDetectors}$ is initialized using the function $init\_map(kmeans\_parameters, kde\_parameters,\\ lof\_parameters)$ (Algorithm~\ref{alg:init_map}).

\begin{algorithm}
\caption{Initialize Map of Models}\label{int_map}
\begin{algorithmic}[1]
\Function{init\_map}{$kmeans\_parameters, kde\_pa-rameters, lo-f\_parameters$}
    \State $\text{Map} \langle \text{String}, \text{Model}*\rangle \ \text{map\_classes}$
    \State $\text{map\_classes}["kde"] \gets \text{new kde} (kde\_parameters)$
    \State $\text{map\_classes}["kmeans"] \gets \text{new kmeans} (kmeans\_parameters)$
    \State $\text{map\_classes}["lof"] \gets \text{new lof} (lof\_parameters)$
    \State \textbf{return} $\text{map\_classes}$
\EndFunction
\end{algorithmic}
\label{alg:init_map}
\end{algorithm}

ASTD \textsf{detectors} respects the structure presented in Section~\ref{section1}, except that in order to modularise the specification we use the operator Call, which calls the ASTD \textsf{DetectorInstance} containing the actions allowing the training and the detection by each model. It has two sub-ASTDs \textsf{training} and \textsf{detection} which are of type automaton having both a single state which is initial and final with a loop transition labeled by the same event \textsf{e(userId, ?eventDate: string, ?eventId: string)} and with the following actions~:
\begin{itemize}
    \item $mapDetectors[d]${}$\rightarrow${}$ fit\_partial(data)$ at the \textsf{training} ASTD which launches the computation of the three learning models each time there is enough data in the current window.
    \item $mapDetectors[d]${}$\rightarrow${}$score\_partial(eventDate, event_Id,scores)$ in the \textsf{detection} ASTD , which populates the scores vector with predictions from each model, while adhering to the discrimination criteria set for each of the models.
\end{itemize}


After having obtained the score of each model, we perform a Majority Voting in the  \textsf{majorityVote} ASTD by the action \textsf{Code::majorityVote(scores, alerts, eventId,eventDate)}, which scans \textsf{scores} and in the case that more than 50\% are positive (of value 1), it adds the event data in \textsf{alerts}, as shown in the algorithm~\ref{maj_vote}. 

\begin{algorithm}
\caption{majorityVote: Majority Vote Algorithm}\label{maj_vote}
\begin{algorithmic}[1]
\Procedure{MajorityVote}{$\text{scores}, \text{alerts}, \text{eventId}, \text{eventDate}$}
    \If{$\text{scores.size()} \neq 0$}
        \State $count \gets 0$
        \For{$i \gets 0$ \textbf{to} $\text{scores.size()} - 1$}
            \If{$\text{scores}[i] = 1$}
                \State $count \gets count + 1$
            \EndIf
        \EndFor
        \If{$\text{count} > \left\lfloor\frac{\text{labels.size()}}{2}\right\rfloor$}
            \State $\text{alerts.push\_back}(\langle \text{eventId}, \text{count}, \text{eventDate} \rangle)$
            \State \textbf{print} $\text{eventId}$ is malicious
        \EndIf
        \State $\text{scores.clear()}$
    \EndIf
\EndProcedure
\end{algorithmic}
\end{algorithm}

\begin{figure}[h!]
    \centerline{\includegraphics[width = 5cm,height = 3cm]{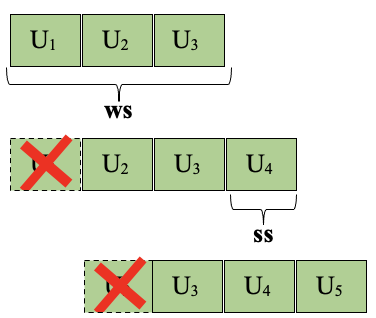}}
    \caption{Sliding window}
    \label{slide_win}
\end{figure}

\paragraph{The method for renewing training data} To apply these models to data streams they are integrated in a sliding window. We have defined three distinct types of windows, to capture relevant information for anomaly detection in various applications, each with varying window sizes. Specifically, we have timestamp-based windows that are categorized based on the number of days or weeks, where each event is associated with a unique day or week number defined by YYYYDDD or YYYYWW, respectively; where YYYY denotes the year, DDD denotes the day's number, and WW denotes the week’s number. We refer to these values as \emph{periods}. Additionally, we have a sequence-based window type whose size is determined by the number of events. In all cases, the window's initialization involves the following three parameters:
\begin{itemize}
    \item $window\_size$: the number of days, weeks, or events the window covers.
    \item $sliding\_size$: the number of days, weeks, or events the window moves.
    \item $type$: 'day', 'week', or 'instance'.
\end{itemize}
Window sliding is shown in figure~\ref{slide_win}, and depends on two parameters: $window\_size$ (\textsf{ws}) and $sliding\_size$ (\textsf{ss}). $Window\_size$ consists of three units, representing the window size, and $sliding\_size$ consists of one unit, which determines the number of units by which the window moves; data associated with old units is deleted. The window moves when we obtain the necessary data to complete the $sliding\_size$, at which point we update the window and delete the data from the previous window's old units

\paragraph{The training and detection by each model} occurs as follows:
\begin{itemize}
    \item \textsf{K-means}: Throughout the training process, our objective is to identify the clusters within the data of the current time window. We optimize the number of clusters, denoted as 'k', by evaluating the silhouette coefficient, a measure of cluster quality. In addition to identifying the clusters, we also compute and store the standard deviation and mean values for each cluster. During the detection phase, our system identifies anomalies by a two-step process. First, we determine the cluster that is closest to the input minute, and then we calculate the z-score. If the computed z-score exceeds a threshold defined by 'kmeans\_parameter', we classify it as an anomaly.
   \item \textsf{KDE}: The training involves modeling the probability density of a user's activity over the 24 hours of the day based on the data contained in the current window. The percentile of the probability densities of the training data is calculated according to kde\_parameter, which represents the detection threshold. Then, when a new event is received, the time of its occurrence is calculated. If the probability density associated with this time is below the threshold, the event is assigned a value of 1, indicating that the event is considered an anomaly.
    \item \textsf{LOF}: We use the algorithm from the sklearn library, choosing cosine as the metric. Before providing the data to the training model, we convert it into Cartesian space to ensure compatibility with the chosen metric. The percentile of the LOF scores of the training data is calculated according to lof\_parameter, which represents the detection threshold. When a new event is received, we compare its score with the threshold. If the score exceeds the threshold, the data is considered abnormal and is assigned a value of 1.
\end{itemize}

\section{Experiment}\label{experiment}
The initial goal of this case study was to detect user activities occurring at unusual times within the activity logs of various Microsoft Office 365 services~\cite{b13}. However, since there is no ground truth available for this type of application, we will apply the case study to a dataset from CERT Insider Threat version 4.2~\cite{b18}, which simulates the activity of 1,000 employees, 70 of whom are malicious according to three malicious scenarios. The dataset that will be used is logon.csv, which contains user IDs, logon and logoff dates, and the PC on which the activity was performed. We will focus on the anomalies associated with the first scenario, which identifies users who logged in after working hours to upload data to wikileaks.org.

Although we will concentrate on a subset of the available information, our main interest in this application lies in the detection rate. We will also examine the effect of combining models through majority voting and the impact of the data renewal method. We are not concerned with false positives since we are not using the complete dataset, and an event occurring outside regular working hours may be normal, considering the role of the employee who performed it, as well as the nature of the PC (shared or private).

First, we convert each line of the logon.csv file into an event in the form of e(userId, date, eventId), and then we provide these as input to the executable C++ code that translates the ASTD specification.

The evaluation is performed using the detection rate (DR) for different models. The detection rate (DR) is defined as follows:

\[
DR = \frac{\text{True Positive}}{\text{True Positive} + \text{False Negative}}
\]

\begin{table}[h!]
\centering
\begin{tabular}{|l|l|lclclcl|}
\hline
\begin{tabular}[c]{@{}l@{}}Thresholds\\ (kde, lof, kmeans)\end{tabular} &
   &
  \multicolumn{7}{c|}{(1.5, 0.5, 95)} \\ \hline
\begin{tabular}[c]{@{}l@{}}window\\ (window\_size,\\ sliding\_size,\\ type)\end{tabular} &
   &
  \multicolumn{1}{c|}{kde} &
  \multicolumn{2}{c|}{lof} &
  \multicolumn{2}{c|}{kmeans} &
  \multicolumn{2}{c|}{\begin{tabular}[c]{@{}c@{}}combined\\ models\end{tabular}} \\ \hline
  \multirow{3}{*}{(10,5,week)} &
  DR &
  \multicolumn{1}{c|}{0.89} &
  \multicolumn{2}{c|}{0.89} &
  \multicolumn{2}{c|}{0.94} &
  \multicolumn{2}{c|}{0.92} \\ \cline{2-9} 
 &
 
  Number of alerts &
  \multicolumn{7}{c|}{109060} \\ \cline{2-9} 
  \hline
  \multirow{3}{*}{(10,0,week)} &
  DR &
  \multicolumn{1}{c|}{0.73} &
  \multicolumn{2}{c|}{0.19} &
  \multicolumn{2}{c|}{0.54} &
  \multicolumn{2}{c|}{0.48} \\ \cline{2-9} 
 &
 
  Number of alerts &
  \multicolumn{7}{c|}{175995} \\ \cline{2-9} 
  \hline
  \multirow{3}{*}{(100,50,instance)} &
  DR &
  \multicolumn{1}{c|}{0.60} &
  \multicolumn{2}{c|}{0.56} &
  \multicolumn{2}{c|}{0.84} &
  \multicolumn{2}{c|}{0.76} \\ \cline{2-9} 
 &

  Number of alerts &
  \multicolumn{7}{c|}{100948} \\ \cline{2-9} 
  \hline
\end{tabular}
\caption{Performance metrics for an anomaly detection configuration using KDE, LOF, and KMeans models with different window settings.}
\label{tab1}
\end{table}
The table~\ref{tab1}, highlights the performance metrics of different anomaly detection configurations, revealing notable improvements when combining KDE, LOF, and KMeans models using a majority voting method. Across all scenarios, the combined models show enhanced detection rates (DR), indicating that leveraging the strengths of multiple models results in more robust and accurate anomaly detection. This ensemble approach reduces the likelihood of missing true anomalies while maintaining a high overall performance.

Comparing the different window configurations, we observe that the first case (10,5, weeks), with a window size of 10 weeks, a sliding size of 5 weeks, performs the best. This setup achieves the highest DR  value across all models, demonstrating its effectiveness in detecting anomalies. The large number of alerts generated in this configuration indicates the model's high sensitivity. In contrast, the second case (10,0, week), which lacks a sliding window, shows significantly lower performance metrics. The absence of overlapping windows reduces the model's ability to renew data, leading to decreased detection rates values, although it generates more alerts, potentially increasing false positives. The third case (100,50,instance) uses an instance-based sliding window, resulting in moderate performance. While the combined models still outperform individual ones, the overall DR values are lower than in the first case, and the number of alerts is the lowest, suggesting fewer false positives but a risk of missing true anomalies.



The use of a sliding window improves anomaly detection by enabling continuous data renewal, which helps maintain high accuracy values. The ASTD specification ensures consistent performance across users by standardizing data size.

Choosing optimal parameters such as window size and sliding size is crucial for effective anomaly detection and minimizing false alerts. Optimal parameter selection enhances model accuracy and reliability across various scenarios.

\section{Discussion}\label{discussion}
The ASTD specification in Section~\ref{section2} uses the quantified interleave operator, which provides processing independence for each user and allows separation of the variables common to all users from those associated with each user. The common variables are defined at the level of the quantified interleave, while the user-specific variables are declared below the quantified interleave in the specification hierarchy. These variables can be accessed by their names in the specification, and the cASTD compiler forwards them to the associated \textsf{userId} instance. Data renewal is performed using a sliding window approach, which requires three specified parameters: $window\_size$, $sliding\_size$, and $type$. The management of data and launching of recomputation of learning models for each user are dependent on these parameters.

The ASTD specification in Section~\ref{section2} illustrates the utility of the Quantified Flow operator, which preserves the modularity and extensibility of the specification, while effectively leveraging object-oriented principles. It also enables the execution of the three learning models while maintaining the functioning of each abstract. The combination of learning models is achieved through Majority Voting, but other combination techniques can be employed by modifying the action at the left sub-ASTD of the ASTD \textsf{combination}.

The ASTD language provides a framework for better structuring the code by its operators. It enables us to determine the different components of the system, in our case: user, learning models, and window, as well as their interrelationships. This promotes the adoption of a robust development approach. However, the C++ language, which is employed at the level of actions in an ASTD specification, does not currently provide a comprehensive set of machine-learning libraries. This limitation could be addressed by integrating Python code that handles these tasks.

It's worth noting that the ASTD specification presented here is not limited to the specific anomaly detection methods described. Instead, it can be easily adapted to accommodate various other anomaly detection techniques by simply modifying the ASTD components specific to the chosen method. This flexibility underscores the generative power of the ASTD language.

Additionally, the object-oriented architecture of the classes representing the learning models plays a pivotal role in abstracting the specific behavior of each model. This architectural choice harmonizes seamlessly with the ASTD framework within a Quantified Flow. As such, our contribution extends beyond a practical implementation and serves as a specification pattern, emphasizing the language's capacity to abstract and modularize complex systems.

The ASTD language, through its visual approach, provides a detailed view of the various stages of the pipeline, thus allowing for a better understanding and maintenance of the system. This becomes more apparent when using the eASTD editor of the language, where for each component of the specification, one can see its various properties and also assign comments describing its function in the overall system.

One of the major properties of the ASTD language lies in the modularity it brings to the development of detection systems. This modular approach not only facilitates the initial development of the system but also its subsequent evolution. A designer can make targeted modifications without compromising the overall integrity of the system.

Another important aspect of the ASTD language lies in the scheduling of the features of the detection system. The language's operators play a central role in this task, enabling smooth and efficient process management. By entrusting scheduling to these operators, the ASTD language significantly reduces development effort. Designers can focus on business logic, leaving operators to handle the coordination of different stages of the system.

The drawback of this method lies in the fact that it requires an understanding of the functioning of each of the ASTD language operators. Indeed, although the clear visualization and modularity offered by the language facilitate system design and maintenance, dependence on operators can pose a challenge for developers less familiar with them. Note that the purpose of this experiment is not to evaluate the accuracy of the produced detection model.  This is a separate issue that is orthogonal to the objective of this paper, which is to streamline the construction of models. 

\section{Conclusion}\label{conclusion}
In this study, we use the ASTD language for anomaly detection in data logs. Our focus centered on the sliding window technique for continuous learning in data streams, coupled with updating learning models upon the completion of each window to maintain accurate detection and align with current data trends. Additionally, we emphasized the significance of employing methods for combining learning models, especially in the context of unsupervised learning, which is commonly used for data streams. To facilitate this, we extended the ASTD language with a new operator, the Quantified Flow, which enables the seamless combination of learning models while preserving the functioning of each of them in an abstract manner. Therefore, our contribution extends beyond a mere implementation and serves as a specification pattern, highlighting the language's capacity to abstract and modularize anomaly detection systems. In conclusion, the ASTD language provides a unique approach to developing data flow anomaly detection systems, grounded in the combination of processes through the graphical representation of the language operators. This simplifies the design task for developers, who can focus primarily on defining the functional operations that constitute the system.

\end{document}